# Inverting normative city theories and computational urban models: Towards a coexistence with urban data streams[1]

Short title: Inverting normative city theories and computational urban models

Diana Alvarez-Marin[2], Vahid Moosavi[3]

# Abstract

Two unavoidable processes punctuate our century: The unprecedented urbanisation of our planet (United Nations, 2014) and the spread of ubiquitous computing (Weiser, 1991) and urban data streams. This process of urbanisation corresponds with the process of digitalisation of urban life: while urbanisation acts on a physical infrastructural level, the digital develops as a kind of *metastructure above the infrastructure*. This metastructural level offers a flexible framework through which information is continuously and operatively being *symbolized.* Today, Information technology and the availability of abundant urban data streams could be considered as forerunners of our time, having unprecedented impacts comparable to the ones brought by the steam engine at the dawn of industrialisation and the electrification of cities. It is therefore no longer conceivable to think of the physical structure of the city without including its digital counterpart.

Against this background, we will explore the role of computational power and information technologies as dominant factors in the formation of computational urban models and normative city theories. We will show how these models and theories, emerging mainly during the 19th and 20th centuries, present leaping correspondences with more ancient conceptions of the city, when observed from a meta-level or episteme (Foucault, 2002) approach. First, and for the sake of clarity, we will deal with some methodological

---

elucidations around the concepts of theory, model and episteme, and how we will refer conceptually to these terms throughout this paper. Secondly, we will review these evolving technological and computational *levels of abstraction* and their influence on the different conceptions of the city. Thirdly, we will develop the hypothesis of a *conceptual gap,* between our current technological capacity – grounded on the abundance and availability of urban data streams – and the state of the art in urban modelling and city theory. Lastly, building upon Foucault's concept of *episteme* (Foucault, 1970) and *genealogy* (Foucault, 1977b), we will explore this gap by speculating around the possibility of an inversion in computational urban modelling and city theory. And above all, we will question the terms in which we can think of the city, in an age where the world can be virtually conceived as fully urban, and the continuity and abundance of urban data streams giving account of it can be taken for granted. How are we articulating the phenomena we call city on top of this generic common ground?

# Introduction: Model, theory, genealogy

The Oxford English Dictionary defines theory as *a supposition or a system of ideas intended to explain something, especially one based on general principles independent of the thing to be explained*, and a model as *something such as a system that can be copied* or a *simple description of a system, used for explaining how something works or calculating what might happen, etc*. Under this light, theory places itself on a metalevel in relation to the model: while the former explains a certain condition through general principles, the latter is an established system playing within a certain framework. Consequently, theories are structural for any modelling; there cannot be any model without a theory, even if this theory has not been made explicit (Wassermann, 2011).

However, on his more recent critique on models, Batty points out that in cities *any simplification at all is too great a simplification and we loose the essence of what we are interested in.* (Batty, 2015). This argument raises a paradox that this paper would like to tackle: in spite of the abundance of urban data streams and the rising scale and complexity of the urban as a global phenomenon, the question of the city is still being addressed today in simplifying and idealizing terms. The problem addressed in this paper is therefore not the one of urban theories and models per see, but rather the one of the *apparatuses of order* that condition them. As we have exposed previously, we could assume that theories constitute an encapsulating metalevel for populations of models; however, in order to address a problem of *order* around the extensive accumulation of discourses about the city, in terms of their differences and similarities, a more abstract meta level -- encapsulating yet transgressing description, modelling and theory -- needs to be addressed.

Michel Foucault extensively explored the concepts of transgression (Foucault, 1977a), *apparatuses of order* (Foucault, 1980b), and the history of madness and sexuality (Foucault, 1979). Yet, Foucault does not write about history in a standard sense. His *history* is one of disruption rather than progression, one of *epistemological breaks* rather than steadiness. For Foucault *history* does not refer to a *living continuity* or a sequence of causal relations or evolutionary process based on a linear vector of time, but appears rather as a system of shifting relations of power in the discourses of knowledge. Identifying these relations as an *archaeology,* he excavates subjugated forms of knowledge out of more established ideologies by a process of historical differentiation. Foucault's questions are directed instead towards the types of series that need to be established between events or the criteria of periodization that should be adopted for each of them (Foucault, 2012). He goes even further by questioning the grounds on which these forms of excluded knowledge can be considered as true or false, by a method that he calls *genealogy,* against the claims of a unitary or absolute theory that would hierarchize them. For Foucault, a genealogy should be seen as a kind of *attempt to emancipate historical knowledge from that*

*subjection, to render them (...) capable of opposition and struggle against the coercion of a theoretical, unitary, formal and scientific discourse* (Foucault, 2012).

Hence, in order to explore the hypothesis of a *conceptual gap,* between our current technological capacities and the current state of the art in urban modelling and city theory, we will build up on Foucault's genealogy, as a method which is concurrently multidimensional (outside of linear causal historical relations), operational (able to encapsulate any urban theory or model, existent or yet to exist) and at the same time epistemic (focusing on the conditions of possibility of forms of knowledge or ways of articulating the idea of city, rather than established structures of urban knowledge). By no means, we see this stance as a claim for "The End of Theory" (Anderson, 2008) but rather as an evocation towards the possibility of *coexistence* of models and theories.

As representational idealised models reach a limit when facing cities as complex phenomena, we will alternatively abstain from postulating the *why* and the *how* or any kind of underlying mechanism in a given urban condition, as this would represent the perpetuation of the highlighted paradox. Instead, like on Foucault's genealogy, we will focus on developing a unifying perspective for the assessment of different modelling approaches, while preserving all their diversities, to which we will refer here as "pre-specific (Bühlmann and Wiedmer, 2008; Moosavi, 2015). Finally, we would like to deliberately remind and adhere to Foucault's intention in "The Order of Things" (Foucault, 2002):

I am not concerned, therefore, to describe the progress of knowledge towards an objectivity in which today's science can finally be recognized; what I am attempting to bring to light is the epistemological field, The episteme in which knowledge, envisaged apart from all criteria having reference to its rational value or to its objective forms, grounds its positivity and thereby manifests a history which is not that of its growing perfection, but rather that of its conditions of possibility (…)

# Attempt to a genealogy of the city

In the "City of Theory" (Hall, 1988), Peter Hall points out the major transition that the fields of urban planning and design were exposed to during the 1950's. During this decade computers, although still rare and expensive, provide the framework on which systems analysts develop a new paradigm in city planning relying on models and accurate methodologies rather than *old fashioned* architectural blueprints. During the 1970s', with the increase in computational power and the advent of the "Systems Revolution" traditional models become more elaborate and detailed, while urban planners and city theorists develop tripartite normative theories describing successive ages in civilisation (Shane, 2005), classifying the diversity and width of existent urban models within a larger encapsulating frame.

There are several comprehensive literature surveys that review both, the fields of urban modelling and city theories from a chronological perspective such as Batty's (2009) or Wilson's divided to spatial, temporal and functional scales (2012). Amongst the most prominent examples in city theory we could list Kevin Lynch's Normative City Theory (Lynch, 1984), Cedric Price's Egg Diagrams (Shane, 2005) and Spiro Kostof's Three City Models (Kostof, 1991). (See Fig. 1). More recently, Brenner and Schmidt (2015) have posed the question of a new epistemology of the urban, by investigating the categories and methods of urban analysis by which the contemporary urban condition should be examined. Even tough we adhere to idea that a universal or naturalistic form of urban settlement is not a viable approach, we believe a new approach doesn't require to reinvent key categories of analysis in relation to ongoing processes of historical change (Brenner, 2015), but demands rather to get rid of them. The urban phenomena is a complex one, and any attempt to define it in terms of dimensions or parameters becomes arbitrary as a pre-selection always needs to be done by the modeler or observer in question. This critique stands on a conceptual level, as it applies to urban theory in as much as it does to computer urban models.

Placing such an emphasis on the new categories and dimensions of urban analysis would suggest for us the establishment of a new theory of the urban. This exactly the point where we consider a genealogy of the urban, rather than a theory, can provide a more abstract approach to a setup in which data, not infrastructures, are subtract. While a theory exposes explicitly how 'things work' or which kind of methodology to follow, a genealogy remains implicit by offering a constellation of indexes to be articulated as "space of the possible", it is a cloud rather than a structure.

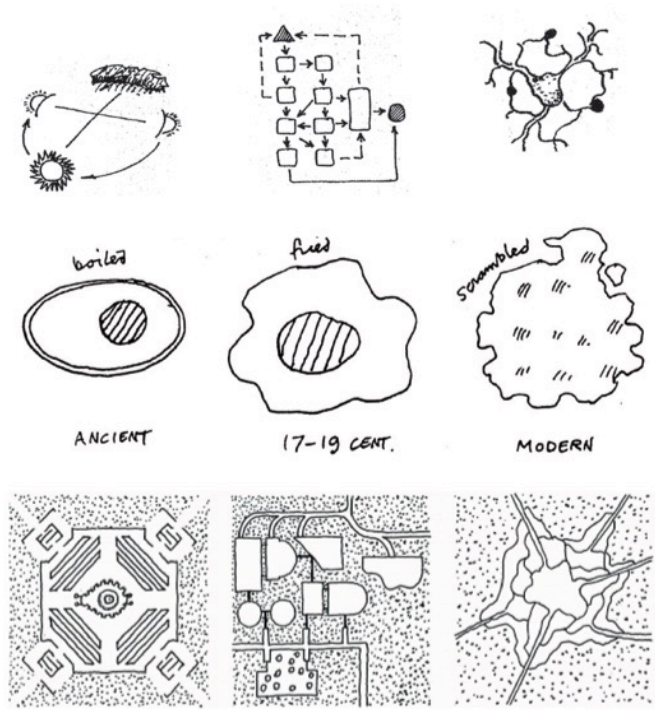

**Figure 1.** Tripartite urban theories

Grahame Shane's Recombinant Urbanism (Shane, 2005) has approached the wide population of urban models from a more abstract perspective, by contextualising urban concepts, often taken for granted, like specific morphologies or systemic metaphors. This gesture towards discussing epistemic shifts *indexically,* in relation to wider cultural and technological contexts, animates our interest. How to build a genealogy of the city where the digital is contemplated as an integral part of it, to the same extent that physical infrastructures and electricity have been considered its main subtracts throughout industrialisation and the 20th century?

Today it would seem almost common sense to affirm that information underpins everything in the universe (Gleick, 2011). According to Serres, information has always been here, *emitted, received, stored and processed* (Serres, 2013), decoupled from time and place and across cultures.

Levels of abstraction in technologies of information are constituted by the ways in which we master this four-fold activity, giving account of our abilities to deal with complexifying urban environments and increasing populations. As pointed out by Townsend (2013), the relation between cities and information technology began in the ancient world. As cities become larger and more complex, so do information technologies, and inversely as the latter provide *space* for more abstract operations cities embrace opportunities for expanding and engendering new lifestyles. Six thousand years ago, the first settlements emerged amid the irrigated fields of the Middle East and served as physical hubs for social networks devoted to commerce, worship, and government. As wealth and culture flourished, writing was invented to keep tabs on all of the transactions, rituals and rulings. It was the world's first information technology (Townsend, 2013). In the nineteenth century, the railroad and the telegraph contracted distances and played a major role in the expansion and development of the industrial city, certainly not merely led by mechanic physical power but clearly also by information technologies. In the last century, we invented a machine that allows for the abstraction and acceleration of the *emission, reception, storage and treatment of information,* a strange machine with four universal rules, a universal machine (Serres, 2013). Computers, these universal machines, are the ultimate mental amplifiers – computers can amplify any intellectual activity we can imagine (Evans, 2011).

From this perspective, and in order to ponder seriously the latest relevance that has been given to information and "Big Data" in the fields of urban modelling and city theories, we will integrate in a proposed genealogy the role of computational power along a capital notion in Computer Science: the concept of *abstraction*. By mapping a landscape of singular marks, rather than constituting an ontological system of categorization, we will argue on four *regions* or *levels of abstraction* (See Table 1) inside a spectrum of cities and information, *entering, knowing, connecting and learning*.

**Table 1.** Main concepts and structure in the four levels of abstraction examined throughout this paper.

| ENTERING | KNOWING | CONNECTING | LEARNING |
|---|---|---|---|
| DESCRIPTION | COMPUTATION | CONNECTIVITY | DATA STREAMS |
| NATURAL MODELS | AGGREGATED MODELS | DISAGGREGATED MODELS | LEARNING MODELS |
| GEOMETRY | LOGIC | TOPOLOGY | PROBABILITY |
| CORPUS | MODULE | PARAMETERS | INDEXES |
| DIVINE LAWS | ALGORITHMS | SYSTEMS | CODING LITERACY |
| COSMOS | MACHINES | ORGANISMS | PATTERNS |

# Level 1: Entering

Cain was the first city builder, substituting God's Eden for his own. For Ellul (1970), the importance of this affirmation is not the veracity of this fact, but rather its significance, which gives a human view of the divine. Under this light, acts of foundation come as the development of a human consciousness of being aside from nature, yet they imitate a celestial archetype. For Elliade (1954), cities re-enact a reality that transcends them, for they commemorate a mythical act of foundation or the very creation of the world, becoming hence a hierophany or a sensorial manifestation of the sacred. We will proceed now to explore how these foundational acts or rituals manifest not only in archetypical but also in contemporary urban and computational models.

## 1.1 Description and natural models

Before the emergence of concepts such as programmability or even mechanization, modelling is based on specific descriptions of reality. Descriptive theories in as much as natural models, are concerned with providing a timeless description of nature. Descriptive thinking precedes algorithmic thinking, as its focus is not on the repeatability of a process – a sequence of steps – leading to a certain solution, but rather on a top-down holistic view to a given problem. In computer science, we could relate descriptions to functions or equations – universal laws – or a sort of primitive procedures, mapping statically from inputs to outputs, where for each valid input to the function there is exactly one associated output (Evans, 2011).

On "A Short Course for Model Design", Lowry (1965) argues that descriptive models reveal much about the structure of the urban environment, reducing the apparent complexity of the observed world to coherent and rigorous languages of mathematical relations. Drawing on Lowry's scheme, Batty refers to these models as "Iconic", as they visually convey what the real things looks like, as scaled down versions of it (Batty, 2007). This kind of models is mainly grounded on geometry.

Classically there are two general approaches for solving a mathematical model, known as *analytical and numerical approaches.* An *analytical solution* aims at devising how a model will behave under specific circumstances – descriptive theory – delivering a *closed form solution.* Lowry asserts that analytic solutions are only applicable *to models which exhibit very tight logical structures and whose internal functional relationships are uncomplicated by nonlinearities and discontinuities.* These models are solved by analysis by putting into direct relationship the output variables and the input variables, in a way in which intervening variables drop out of the reduced form equations (Lowry, 1965). As a result, in this setup only one solution is possible to a given problem.

## 1.2 Geometry and corpus

Descriptive approaches are therefore easily translatable into specific geometries, governed by a set of equations or *divine principles* that can be translated into non-volatile, established and controllable conditions. In this line of thought, Lynch's cosmic city governs the finite boundaries of its domain within explicitly identified laws and accurate proportions. (Lynch, 1984). Concurrently, Cedric Price's "Boiled Egg" model is defined purely in geometric terms: dense, fixed in concentric rings of development and still encased in a shell, alluding to the organisational structure of a nucleus within a perimeter wall.

Economical-spatial models introduced in the mid 50's, inspired by German Location Theory (Von Thünen, 1826) and Ricardo's economic theories (Ricardo and Li, 1819), place economy at the centre of a whole system of geometrical relations. Location Theory addresses the question of economic activities and their location, based on the assumption of the pre-existence of ideal distances manifested in specific functions or universal laws. In this line of thought, Von Thünen and Schumacher-Zarchlin's mathematical theory of productivity (1875) defines a centralised hierarchy between markets, production and distance, while Weber and Pick applies the same principles on industry (1909) by devising a rule that identifies optimal location for manufacturing. Coupled with the development of logical positivism, these models flourish as a means for the instauration of general hypothesis about urban spatial distribution that could be consecutively tested.

## 1.3 Divine laws and cosmos

Elliade places the divine as archetype and inspiration as *it is only for their gods that men exert themselves so extravagantly* (1954), reaffirming the idea of the city as cultural cosmos or miniature representation of the universe. In this line of thought, Rykwert argues that the geometrical layout of a town embeds the cosmos itself through the manifestation of specifically cultural divine laws (Rykwert, 1954). He stresses further the primacy of these acts of settlement by affirming the concept of town, not as rational or functional phenomena, but rather as a conceptualization of rituals, myths and crafts. These myths situate us in a universe within which we can recognize ourselves, reaffirming the idea of the city as cultural cosmos or miniature representation of the universe (Rykwert, 1988). Rykwert argues that the geometrical layout of a town embeds the cosmos itself through the manifestation of specifically cultural divine laws.

In this setup cities can be seen as micro-cosmos, all parts bonded into perfectly ordered wholes, which do not posses urban alphabets nor combinatorial logics but rather all pervading, divine ordering principles. Their centres are not places of maximum concentration nor economic interest but rather places of identification, the ones that distinguish themselves from other places.

We could consider this approach as monist, as one and only one set of principles explains simultaneously both natural and the cultural realms. Materialist Monism (Curd and McKirahan, 2011) aims at providing an explanation of the physical world by means of a single element (Curd, 1996), where the one can be identified with the whole cosmos and inversely. Already in 1928, Sorokin (1928) argued the emerging field of Social Physics to be based on Ancient Greek Materialistic Monism, as *physical and social phenomena [were] mere variations of natural phenomena* (Barnes and Wilson, 2014), assuming that both human and non-human worlds could be explained by a single unified theory coming from Physics and the description of nature.

More recently MIT Media Lab, empowered by the advent of Big Data as substrate for a new Social Physics and by the quantification of social relations, has developed a predictive, computational theory of human behaviour (Pentland, 2014). The emerging field of city science (Bettencourt et al., 2007) has been attracting a community of physicists with a classical Newtonian approach, asserting the possibility of an absolute universal principle for cities or any kind of human agglomeration, able to prescribe *energy consumption and size, socioeconomic outputs, from economic production to innovation but also crime as proportional to the number of social interactions.*

In spite of relying heavily on computational power, these models are based on analytics; consequently their idea of the city is static and aggregative. Their order exudes a certain automatism of the cosmic where divine principles can be replaced by scientific or economic ones in a timeless way.

## Level 2: Knowing

In the late 17th century, attempts to apply classical mechanics to describe relationships between people and places, under the form of Newton's Laws of Motion, become common rule. Consistent with Location Theory, equations from natural science and classical physics applied to large population numbers come in force in the mid 1940's. According to Harris (1964), Steward and Zipft pursued the arguments of Social Physics, by transposing generalisations from the realm of classical Physics to Social Systems, which became later on popular during the 1960's under the influx of the increasing computational power.
In 1936-37 Turing (1936) creates the concept of Turing Machine, a virtual engine provided with a certain fixed function, intended to investigate the limits of what can be computable through logical and numerical instructions. Turing's machine introduces a time vector: it is capable of giving account of current state, follow rules matching this state and determine the next step to follow. The number of states this machine can present is however limited, as Turing argues that a human would only be able to keep track of a few amount of states at the time (Evans, 2011). With Turing's invention, all the ingredients of the actual digital computer are disposed, however the latter would only be designed in the early 1950s.

## 2.1 Computation and aggregated models

In 1948, Wiener (1961) installs Cybernetics as the science of "Control and Communication in the Animal and the Machine". Wiener proposes a technique to approach the problems of control and communication between humans, living beings or machines undifferentiatedly, placing human beings not as individuals in charge of their own destiny but as components reacting to predefined functions. The mechanisation of descriptions into algorithms characterizes a shift from an analytical towards a numerical setup. Computers allow for the implementation of numerical methods, as these involve a great number of repetitive, but very simple calculations, and provide solutions for more complex problems. The result is not an equation but rather a long list of numbers that represent a range of possible and approximate solutions.

Numerical methods are executed by consistent systems of logical algorithms or *rational models*, in order to imitate certain real world phenomena such as urban traffic, land use dynamics, economic activities and so on. This hierarchical and consistent set of equations (or rules) is considered as the surrogate generalisation of the phenomena under study and time-based changes are integrated, embodying a shift from *static descriptive models* towards *models in dynamic equilibrium*, or simulations.

Although computational power opened up new possibilities for better understanding of real world phenomena, these computational models became data hungry and their demand for data was higher than what was available for model tuning and validation. The simulation of a target system's behaviour, reached a threshold when dealing with complex, unpredictable and open systems, originated on one side by limitations on calculation power and on the other by their underlying algorithmic representations, producing some scepticism about the applications of computational models to real world problems (Lee,

1973).  We consider this level of abstraction as computational or machinic, and based on the prescription of an ideal form following principles of some rational or moral objectives.

## 2.2 Logic and module

Towards the mid 1950's, the development of integrated circuits dramatically increases the efficiency and speed of computers, leading to a fascination for the quantification and analysis of diverse societal phenomena under the form of mathematical models. The introduction of mainframe simulations and structured programming, also called high-level or procedural languages, shifts the focus from the linear sequences of instructions towards a flow of operations, integrating higher-level constructs such as while loops (iteration) and switch statements (selection) (Scott, 2000), that we will see clearly manifest in concurrent ideas of the city.

For Lynch (1984), the machinic manifests in processes of colonisation, where additive systems of rules allow for uninterrupted urbanisation and ubiquitous power. Modularity consents extension by addition without perturbing a general order where parts can be replaced independently. According to Mumford and Copeland (1961), the Greek democratization of the urban block or module, assigned to a certain type of an established urban vocabulary, provides not only international legibility but allows equally a commodification of the land that denies its cosmic memory. By importing the Greek grid, the Roman city inverts from a political focus on social order and distribution towards an economic emphasis on expansion ex-novo and territorial organisation. In a similar fashion, for Pope (1996), the American land expansion under the regime of the "Jeffersonian Grid" indicates not just democratic allocation, but defines an obvious national destiny, regulating and predefining everything that is and there is to be (Pope, 1996).

## 2.3 Algorithms and machines

On this level of abstraction, the identification of a general law as natural order is no longer sought after but rather its prediction and mechanization. We identify here a shift from a cosmic diagnosis towards mechanical prognosis, where model the aim is to predict its state in time; therefore a causal consequence must be specified. An algorithm is mechanical procedure that can be expressed within a finite amount of steps (in space and time) for calculating a function. Starting from an initial state and initial input, the instructions describe a computation that, when executed, proceeds through a finite number of well-defined successive states, eventually producing output and terminating at a final ending state.

Already in the 15th century Alberti's concept of *lineamenta* (Alberti and Rykwert, 1991), as the outline to be perceived in the design organized within a system of rules and algorithms, introduces a clear distinction between object and design where an *instantiated object* becomes an identical copy of its own design. Based on this process of abstraction, in "Descriptio Urbis Romae" (2000), Alberti describes a method to record information from his survey of Rome using a system of polar coordinates, encoding reality as a list of coordinates linked through a list of instructions or primitive algorithm, so that the resulting map of the city is reproducible at desire, based on his observations. With this discrimination between object and abstract object, Alberti places the cornerstone of modernity.

In computational urban modelling, approaches focused on spatial and urban morphology aim at understanding the spatial structure of cities by examining the patterns of their component parts and the process of their development, based on the search of top-down, macro principles. Under this light, cities appear here as closed systems (or machines) over any given state, specific object or architectural scale. In geography, Lewis Mumford, Peter Hall and Michael Batty influence this approach greatly, while Hillier and Hanson (1984) interprets it as network analytics, based on idea that space can be broken down into a network of components and people's movement defined through universal assumptions on how we make choices. The interaction of these *laws of spatial emergence* and *generic functions* is known as "Space Syntax" (Hillier, 1996).

Cities can be here seen as deterministic: focus is no longer centered on particular cities per see, but rather on a system of iterative algorithms. These models are highly centralized and only use computational capacities in order to simulate a system that can be qualified as aggregate and in *dynamic equilibrium.* Time is manifest through the introduction of states and causal relations, replacing geometric space as ground, yet remains reversible as it considers the city as a closed system.

# Level 3: Connecting

The increase in computational power in conjunction with the new fluctuating urban condition that came during the post-war reconstruction period, leads to the design of more accurate and detailed models, which paradoxically become more difficult to validate, as there is no essential stability in the system they model. A wave of movements and ideologies emerging as a critique of modernism after the Second World War, is clearly manifest in the diagrammatic concepts of relational city, where relations become primary over nodes, and flows over geometry (Smithson, 1957). In architecture and urban planning, Peter and Allison Smithson and the TEAM10, introduce the concepts of cluster, and flow, as a more complex stance to the "geometry of crushing banality" of the Moderns (Smithson, 1957). In their article "Cluster City, a New Shape for the Community", they affirm,

Le Corbusier's dream of a Ville Radieuse was supported by a geometry of crushing banality. (...) What we are after is something more complex, and less geometric. We are more concerned with "flow" rather than with "measure". The general idea which fulfills these requirements is the concept of the Cluster.

Diverse critiques emerge along post-modernist and anti-planning views of urban theory. In "Non-Plan: An Experiment in Freedom" (1969), Peter Hall, Cedric Price, Reyner Banham, and Peter Barker argue against urban planning and promote instead the idea of decentralised, flexible and adaptable local situations. In "Theory and Design of The First Machine Age", Banham (1980) criticizes equally the static and standardized nature of the modernist approach. Banham and Archigram, in a way, predict a shift from archetypal urban machine (computation) to the concept of megastructure (network) where the materiality of a built process is replaced by an actuated evolving territory. Archizoom's Non-Stop city (Branzi, 2006) embodies this idea radically, by conceiving the city as an activated and endless landscape consumed into 100% urbanisation, a complex infrastructural mass providing freedom and ubiquitousness to congregate anytime anywhere. Similarly, Nieuwenhuys (1974) and Friedman's (2000) utopias picture the deployment of a centralised single system as territorial climate, a neutral and pervasive grid capable of accommodating unlimited possibilities. We would like to consider the idea of megastructure as *activated field* or *ubiquitous machine,* equally as a transitional ideological state, focused on processes and relations rather than discrete parts.

New urban models introduce concepts of emergence and dynamic systems *far from equilibrium,* based on analogies or imitations of other systems most of the time biological.. Under this light, Wolman's Urban Metabolism (1965) retraces the city as a biological eco-system updated by networks of information inflows and outflows. Alongside, Forrester's Urban Dynamics (1969), originally developed to improve efficiency in industrial processes, introduces concepts such feedback loops and stock and flows in the idea of city. In the early 2000's, Alexander (2002) and Salingaros (2005) develop a new school of Urban Morphology inspired from emergence and morphogenesis, arguing for the existence of an underlying nature of order (and the universe), as patterns are not enough as to trace the principles to a good built environment. Under this light, the idea of city appears as a computational process alike to cellular growth in a living organism.

## 3.1 Connectivity and disaggregated models

In the 1960's, ARPANET is developed under the direction of the U.S. Advanced Research Projects Agency (ARPA), becoming the network basis for the Internet. Based on the idea of sending information in small units (packets), ARPANET is able to route information packets on different paths and restore them at their destination. The development of the TCP/IP protocols in the 1970s makes it possible to expand the size of the network, into a network of networks.

A second wave of cybernetics emerges in the 1970's embracing the idea of computers fusing together based on a natural order where society, as a network of free and equal individuals linked together in a global system, would be able to find its own natural order. In a similar fashion, urban theories shift from *centralized to decentralized,* while *aggregate models* rapidly turned into *disaggregate,* which no longer possess a central system of equations, but instead individual agents acting based on differentiated internal algorithms and goals (like humans as agents or cells in the land use models), while interacting with the other agents through (designed) communication channels. Therefore, by changing the interaction mechanisms as well as internal behaviours of agents, different emergent properties on top of individual models are expected.

From 1980s, as the concept of computational networks emerges, with more data and more powerful computers, an interest for detail at the micro scale arises, along the development of GIS and raster based representation. Alongside the Internet, mobiles computing and networks of sensors start to spread pervasively, and as a result, the amount of digital data increases dramatically. In the mid 1990s, terms such as "data mining" and "database management" emerge alongside methodologies for the exploration of digital data (mainly structured data) among modellers.

In this setup, the focus shifts from *algorithms* towards *connectivity*, from *computing machines* towards *networks of computing machines*, from a *dynamic closed world in*

*equilibrium* towards a *pulsating and open ended world far from equilibrium.* This level of abstraction offers a new architecture of modular systems with heterogeneous layers, but at the same time presents a problem of integration and complicatedness in its models. (See Lee, 1973) Data starts to emerge – quantitatively rather than conceptually -- but only as a derivative of network collection capacity, being pre-structured by the system or the modeler's parameters of choice.

3.2 Topology and parameters

In the 17th century Gottfried Leibniz envisages the concepts of geometria situs (Greek-Latin for geometry of place) and analysis situs (Greek-Latin for picking apart of place), establishing the base for the development of topology as a field of study out of geometry and set theory. For topology, some problems do not depend on the corpuses or exact shape of the objects involved nor their internal algorithms, but rather on the way they are put together in order to shape an emergent manifold. Networks are topologically arranged, both geometrically, in terms of the location of its nodes, and logically, in terms of how data flows inside the network or network protocols. Once these two conditions can be considered as ground, the focus shifts towards the specification and tuning of systems of algorithms, through parametrisation.

With the microprocessor, smaller computers become more powerful and can be linked together into networks, introducing as well a decentralization of computation from commercial to personal computers. The introduction of Object Oriented Languages allows for the encapsulation of data and methods of manipulating data units called *objects,* interacting with one another in such a way that their methods can be modified without affecting any code that uses them. Deprived of any specificity while yielding many specifications, *these abstract objects* produce populations of instances of one kind, and become the corner stone of Parametricism. We can notice here an abstraction of existing abstractions, as an attempt to enhance coding opportunities (Scott, 2000).

## 3.3 Systems and organisms

The fascinosum around the concepts of decentralization, indeterminacy and systems, prevailing in computational urban modelling, takes a metaphoric dash amongst architects and urban planners, to a great extent captivated by biological transpositions on the city. In the 1960's Cedric Price (Haque, 2007) designs "The Fun Palace", inspired by 18th century egalitarian philosophy and Gordon Pask's idea of underspecified goals (Haque, 2007). For Price the city is unsteady, fluctuating and constantly self-organizing through processes of expansion and retraction. "The Fun Palace" exemplified a time-based architecture of underspecified desires, however encapsulated by a fixed containing frame.

     Venturi (1977) introduces the concept of complexity back into architecture and the city, by claiming the need for a return of symbolicity and multiplicity of meanings, metaphor rather than methodology. For Venturi, architecture and the city are about vividness rather than simplification, *both-and* rather than the modernist *either-or.* Rowe and Koetter (1983), intensify and address the fragmentary condition of the postmodern city in "Collage City" through composite collages of ambiguous patterns (or a Vocabulary), that find together a self-organised order. Similarly, Jacobs (1961) stresses the role of the urban-actor designer in maintaining in balance the dynamic *organised complexity* of the city by mutual feedback between agents.

     Concurrently in computational urban modelling, two main branches of micro-simulation, cellular automata (Tobler, 1979), and multi-agent systems (Waddell and Ulfarsson, 2004), explode the centralised function of an urban system into millions of agents functions. The most well known cellular automata model, Conway's Game of Life (1970) sees its day as an infinite two-dimensional actuated grid where emerging cells were binarily assigned with a dead or alive state, 0 or 1. Interaction with adjacent neighbours following a precise set of rules will define the rest. Multi-agent systems require in the other

hand, the framework of a specific model or an a priori rule set, or grammar, in order to model agents' behaviours and interactions.

This approach reaches its limits when building up increasingly complex and detailed models of cities, facing what is known as "The Curse of Dimensionality" (Bellman, 1961). Bellman claims that by adding more and more parameters from the real phenomena into the model, the demand for computational power increases exponentially. As a result, the efforts of such modelling become either very complicated or very simple and not sufficiently distinct for describing the complexity of urban environments with adequate benefit (Moosavi 2015). Similarly for Batty, the enrichment of models by disaggregation and addition of detailed decision-making processes, leads to a paradox where the more detailed the models the harder they are to validate and unlikely to replicate the actual situation they are aiming at (Batty, 2015).

The idea of City becomes thus dependent on the combination of parameters and their respective allowances. Nature is ultimately analogy, after being symbol and opponent. This setup finds *continuity* (and therefore a deadlock) by increasing the amount of parameters and capitalising in structure and computational power.

## Urban data streams and the inversion of modelling processes

The *levels of abstraction* we have reviewed in the previous chapter, do not cancel each other out but cohabit simultaneously depending the degrees of complexity that are being dealt with, and shift along threshold moments of saturation, when more abstract space becomes necessary. For instance, when there is no way to analytically solve a system of equations, the system must be simulated through numerical computation. Which is the case in the problem of urban land use transport models, where highly centralized computational modelling is replaced with a distributed network of computational agents, as relations between different urban actors become more complex. In other words we could relate this to the transition we have previously described from *description and natural models* and *computation and rational models* towards *networks and disaggregate models.*

However, there is not an absolute model of the city — form of life— that can provide a reference plane of stability, but rather a continuous cohabitation of bodies to think in. Higher *levels of abstraction* allow for the encapsulation of greater complexities, as they take their predecessors as ground to stand on. For instance, with Arithmetic we can consider a specific operation 3*4+7, while Algebra by replacing these specific values by variables, confers them with the capacity of becoming *any number.* In the same way, when we are dealing with computer networks, computational units are given; or when we talk about parametricism and variation of an architectural design, we can assume that the idea or "objectile" (Deleuze, 1993) is a given, while the play takes place on the variations and tuning of their established properties.

# The gap

Since the emergence of *the machine* as urban metaphor, appearing along the development of colonies and the masterplans of the renaissance, the idea of city has been coupled with a *level of abstraction*, where the *why* and the *how* – underlying logics, mechanisms and causal relations – are primary. This approach is not reverted but empowered with the arrival of computational networks in the 70's, along numerical computing and simulation capacities (either centralized or decentralized) that have been developing since the 1950's, giving as a result *the organism*, or an infinity of *interconnected machines*, as new urban metaphor. We believe these previous *levels of abstraction* remain *state of the art* today, in spite of the rapid development of a radically new urban condition, the increase on computational power and the availability of urban data streams. Therefore our hypothesis is the one of a *conceptual gap,* between the possibilities provided by our current technological *level of abstraction* and the *state of the art* in urban modelling and city theory (See Fig. 2). Such an assumption is not new, as similar critiques have been discussed in "Requiem for

Large-scale Models" (Lee, 1973), Belmann's "Curse of Dimensionality" (1961), Batty's critique on over-detailed models (Batty, 2015) or even further, Koolhaas' paradox in "Junkspace" (2006), where *the more powerful the solutions deployed to tame the impasse of urban and architectural instability, the larger becomes the increase in scale of what was being affected.*

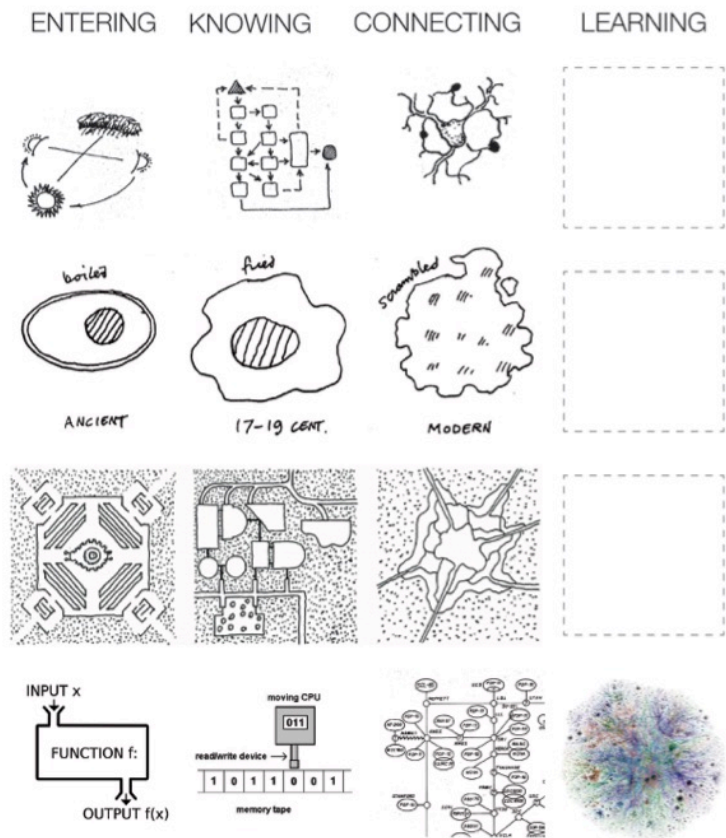

**Figure 2**. Tripartite urban theories and levels of abstraction in computation.

With the concept of "splintering urbanism" Stephan Graham (2001) describes *the fragmentation of the social and material fabric of cities caused by the relegation of urban infrastructure networks and the mobilities they support,* in the field s of contemporary urban studies and urbanism. We consider this scission to be the offspring of decades of fascination with an "urbanism of experts", that developed in parallel with the increase in computational power from the 1950's on. However, at the same time that we claim the return for a transdisciplinary approach to the city that integrates computational urban models and normative city theories, we claim the necessity to "invert" them. The latter means, unlike Graham, our interest is not about infrastructures per see but rather about what can be articulated on top of them and because of them, which is information.

# Level 4: Learning

The bridging of this gap may open up the *conditions of possibility* for an idea of city that integrates its digital makeup as mediation of a concrete urban reality that does not longer require pre-established logics or function to explain its makings. This gesture is not an improvement of a previous level of abstraction, but is rather an inversion of it, as it does not recur to external structures (or infrastructures) but assumes that a certain urban condition when considered contextually and relationally – self referentially – embeds already all its current and possible future estates. Looking at the city from an informational perspective means to go beyond the dichotomy between the theoretical and the empirical, getting out of both a universal concept of the urban and the specificity of local cases. With information we are on a different game, one that cannot be theorized nor structured, for information is neither matter nor energy; it has no form or structure.

## 4.1 Data streams and learning models

Today, with the integration of computation into urban life, we can consider ubiquitous computing as an inversion of virtual reality (Weiser, 1994) and therefore simulation. While the latter is about simulating a computer-generated world, and implies therefore a purely computational power problem, the former is an articulation of contingent human factors manifested as data. In this way, the notion of data appears under a new light and becomes ground, no longer as the result of designed algorithms or experiments supporting a given hypothesis, but rather as raw material emitted through urban activities. These unstructured and continuous flows, or Urban Data Streams, can be considered as a new urban infrastructure.

Next to the challenges these changes bring, we can also see how new areas for research and practice are emerging. To just mention a few, one can refer to Big Data, Data Science, Pervasive Sensing (Hansmann, 2003), Reality Mining (Eagle and Pentland, 2006), Citizen Science (Paulos et al., 2008), Social Network Analysis and Location Based Social Networks (Yuan et al., 2010). The new availability of urban data streams, in conjunction with an increase in computing power, opens a new plateau that challenges the notions of model and simulation, to which we will refer as learning or "pre-specific modelling" (Bühlmann and Wiedmer, 2008; Moosavi, 2015).

With data streams and pre-specific models, the previous establishment of an *abstract object* is no longer necessary, as every real world object can be described *self-referentially* by considering its relations to all other objects constituting its context, or more precisely negatively by considering everything which it is not -- *not all the others* --. In this way, rather than focusing on the systematisation through syntaxes, grammars, and pattern-based schemata, the complexity of everyday urban situations can be encapsulated in a way that *any model* can turn into a *learning model* within a specific condition. Hence, representations of particular objects become increasingly optimal as the amount of data increases. With this in mind, a radical change takes place in the way we think of *data*, shifting from one of empirical validation and verification tool, as is the case with simulation, towards a latent modelling *level of abstraction*.

Today ever growing machine learning algorithms are able to *teach* themselves to perform tasks such as spotting patterns and emergent categories, by crunching huge amounts of data and without being programmed to perform specific tasks originally. In this way the *data deluge* seems to embed its own solution, outside of parametric systems that finds themselves saturated facing the abundance of data. Consequently the focus shifts from the establishment of predefined frames and parametric systems towards unstructured data itself, which embeds any potential system existing and yet to exist, customised to any specific situation.

## 4.2 Probability and indexes

Since the late 1990's two major events mark a radical shift in the ethos of technologies of information. In 1996, Google finds a way of dealing with a trillion indexes planet: the PageRank (Brin and Page, 1998) algorithm, which abstracting from grammars and semantics would be able to represent the probability that a person clicking on certain links would reach any particular page. Later on, in the early 2000's the Web 2.0 (O'Reilly, 2007) or social web sees light as a more interactive and participative platform, which in opposition to its mono-oriented parent version continuously updates within every user intervention. Google articulation of the infinite through indexation is a good example of the potential of unstructured data streams: the content is within the one who asks the question – Google does not produce any content on its own –, while indexes are an evocation around that question, the user picks one, and Google recognizes this answer and shifts its whole world of indexes towards it. This is the way social media creates the non-content index of the content of the world (Hovestadt and Bühlmann, 2014)

Along the rise of the Internet and web base services, scripting languages or *glue code*, intended not for writing applications from scratch but rather for combining already existing components (Ousterhout, 1998), become popular and move towards a co-existence alongside user's activities on mobile devices. We have been progressively shifting from a centralised system traced by infrastructures towards an increasingly decentralised network of applications and data, from *one computer - many people,* to *one person - one computer,* and today we reach a radical inversion with *one person - many devices,* deepened further within their integration into the body and the development of the Internet of Things. Mobile computing opens up an indexical urban landscape that Serres (2013) draws as an unmarked space without distance or magnitudes. For Serres, Cartesian space, marked by coordinates, landmarks, latitudes, longitude or division has disappeared, and shifted instead towards an *indexical space*. Under this light, the space of networks and flows (Castells, 2000) and the generic urban condition (Koolhaas, 1995) are no longer focus but become an acquired ground.

If you ask me my address, (…) this address makes reference to an Euclidean space, Cartesian that refers to given and known reference points. This space is the one in which we have lived and I am going to show that we have abandoned it. It used to be the space of networks (of coordinates, of airways, of highways etc.). As these networks exist since long, we can say that the space of networks is the space of the past (Serres, 2013).

## 4.3 Coding literacy and patterns: the inversion of urban modeling processes

As traditional simulations models construct a logical set up in order to understand complexity under a fixed mechanism, or function, with data streams we are leaving the era of machines. The beauty of computers is that they are not machines. They are abstract machines (Bühlmann, *et. al.* 2015). This means that computers cannot only evoke every known machine, but any machine that we will think of in the future. The ways in which we grasp this *level of abstraction*, beyond the consideration of machines as merely fast calculators or performers of repetitive tasks, can be considered as a new kind of literacy that we will refer to as coding as literacy (Bühlmann, *et. al.* 2015).

We have shown throughout our genealogy, the tight relations woven between our *levels of abstraction* and the ways in which we are able to construct specific ideas of the city (See Fig.3). Taking this into account comes the central question underlying our hypothesis. How is the new coding literacy being integrated to the realms of computational urban modelling and city theory?

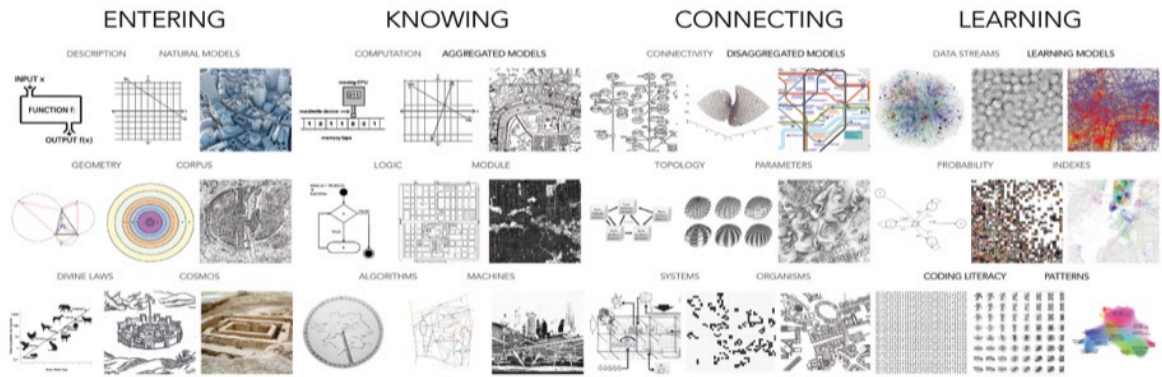

**Figure 3.** Conceptual Matrix on computational and urban levels of abstraction.

An illustrative example of this inversion could be the parallel between the Space Syntax approach to the street network of London (Hillier and Hanson, 1984) and Jing's GPS data visualization of cabs in Beijing (Yuan et al., 2010). Space Syntax is based on criteria of attractiveness and importance of certain street and city segments, consequently the computer only plays the role of calculating machine for the predefined network and data only comes a posteriori as validation means for these structural assumptions. On the other hand, by visualizing a considerable amount of GPS tracks of cabs throughout the day, it is possible to perceive degrees of importance in the urban network and critical points of congestion, without any major assumptions about patterns of human movement. In a similar way, Google Traffic offers a representation of traffic in cities through an aggregation of GPS data emitted by the commuters. Therefore, new types of computational modelling and analysis can be built on top of these data rich platforms rather than classical simulations based on structural or behavioural assumptions, explicated through semantic and syntactic laws and represented by mathematical equations.

Other interesting examples are Eric Fisher's touristic maps of London. Traced without any previous knowledge about the city or even without a pre-established definition of what a street or a block are, just by collecting location and time meta-data from millions of Flickr pictures of London and their correlations, it is possible to trace a map of the city where clusters of touristic attractions appear differentiated from more local areas. This articulation of the city has the capacity of adapting itself in accordance to the shifting interests of local population and visitors alike, in opposition to a more static top-down declaration of these areas. We could also refer to the Livehood Project (Cranshaw et al., 2012) that presents a methodology for identification of land-use patterns in cities, using data from social media and machine learning clustering methods. This approach is orthogonal to classical land use models, where sets of rules of mobility and urban forms are presumed as the underlying mechanisms for defining a neighbourhood.

In a parallel direction, we could refer to diverse mobile applications such as Endomondo, in which running and walking patterns can be used in order to identify emergent walk ways and design better infrastructures for pedestrian flows. By developing appropriate incentive policies based on these observations, city agencies can have an influence and be influenced by the way people are walking in the city, in a way which is neither a top-down modernist planning nor an explicit participatory system, where people's opinion can be biased by the direct involvement in decision making processes.

# Conclusion

We perceive currently a cohabitation of models and theories in the contemporary debate around cities and information, which reflect an interesting overlapping of the levels of abstraction we have described throughout this paper. On one side we recognize *entering* approaches, namely with the emerging fields of Social Physics, which focus on predictive and computational theories of human behaviour (Pentland, 2014), and of city science (Bettencourt et al., 2007). The latter has been attracting a community of physicists with a classical Newtonian approach, asserting the possibility of an absolute universal principle for cities or any kind of human agglomeration, able to prescribe energy consumption and size, while decoupling from any form of cultural contextualisation. This approach asserts consequently that cities are purely definable in terms of fixed universal laws, stressing therefore deterministic ideas of the city and the human, where natural conditions override cultural ones.

Concurrently, on the second level, which we identify as *knowing,* pre-defined algorithms become primary over data. Mostly well known and merchandised under the larger flag of "Smart City", these models are based on the transposition of enterprise managerial processes onto the urban realm and embody private economic interests. Supporters of this approach see integrated digital infrastructures for collecting and processing data as a *second electrification* for cities. Defining cities merely in terms of their infrastructural makeup, may it be physical or digital; this approach presents as ultimate purpose the optimization and performance of the city as finite and controllable *urban object*. More computationally hungry and complex than the *entering* approaches, and its social physics models, machinic models are composed by groupings of rules, translated into systems of algorithms or machines. Oppositely to the previous level of abstraction, these rules are not natural or divine, but dictated by administrative and economic powers (interests).

Alongside, we identify a third approach or bottom-up view of the Smart City, led by hackers, activists and startups. We will relate these views to a *connecting* approach to the city, as networks and relations between agents or machines become primary, and disaggregated models overcome the limits of previously centralized ones. In this line of thought, as Ratti and Townsend put it on the "The Social Nexus" (Ratti and Townsend, 2011)., the best way to harness a city's potential for creativity and innovation is to jack people into the network and get out of the way. Inspired from Jane Jacobs view of city neighbourhoods as organs of self-governance as described in "The Death and Life of Great American Cities" (Jacobs, 1961), this view pre-assumes horizontal structures in city planning would work better than vertical structures. During the last years, this approach has emerged more concretely with the development of civic apps in the early 2010's and continues growing today with the rapid development of digital economy through platforms such as Airbnb, Uber, and crowd funding services such as Kickstater and Indiego. However and paradoxically, in spite of promising a more productive way forward by integrating formerly ignored citizen needs and promoting entrepreneurship, these platforms rely on some kind of centralised corporate states, which provide an infrastructural setup for connectivity. In addition to underlying socio-economical unevenness of its data sets, the *connecting* approach takes perhaps too jauntily for granted the idea that social networks are fundamentally democratic.

*Technology is the answer but what was the question?* It's been almost fifty years since Cedric Price provocatively titled his 1966 lecture with a question that comes back today with tremendous actuality, at a moment where to more complex situations we keep on offering more complex solutions, bigger systems, more parameters, more agents, more computational power. However, Cities have never been about efficiency or optimization. Certainly, 11000 years of urban history barely match with the current standards promoted by rankings of sustainability and liveability. As Greenfield (2013) points out, at the moment, we are only being offered one particular story about the deployment of networked informatics in the urban milieu, and though it is widely predominant in the culture, it only portrays the narrowest sliver of what is possible. Today, with the advent of an increasingly important digital urban make up, we believe architects and urban researchers should look with genuine interest and curiosity at transversal developments in computation and computer science.

This is the reason why we have attempted through our urban genealogy, by taking a step back and looking more abstractly, at illustrating how concurrent informational and technological concepts have a direct influence in the construction of imaginaries and theories about the city. The relation between information technologies and cities has existed since their very inception. However, with the fast development of computers and the exponential increase in computational power followed by the data deluge, we identify an epistemic gap deepening between the level of abstraction these information technologies allow for and our models and theories about the city. When we consider machines purely as fast calculators and confer them with power through simplistically assembled statistics, we are reducing complex entities such as cities to the same kind of deterministic scope, be it centralized or decentralized.

As we have shown on this paper, there's a cohabitation of bodies of thought or epistemes in the contemporary urban realm, yet their scope seems to limit to previous technological capacities. People behave in unexpected and irrational ways that can in no way be predicted by any kind of underlying system, and so are cities, full of paradoxes and contingent occurrences. We believe the crucial question is not the axiomatic solution of problems by establishing a brand data-driven science or theory, nor oppositely the upcoming of a brand new empiricism. While there are new instances that don't seem to fit into the current urban discourse, we see the unfolding of a fourth level of abstraction, which is about *learning* and tends towards a focus in the constitution of problems. "A Quantum City, Mastering the Generic" (Hovestadt et al., 2015) embodies a data intensive or *indexical* idea of the city, which we consider is along this line of thought. Under this light, the perception of patterns should not be considered as answer but rather the key towards new set of questions. This paper was written as an introduction to our body of research, and in our future works we will aim at elucidating the new possibilities that this fourth level on *learning* is opening up.

# References

1. Alberti LB and Rykwert J (1991) *On the Art of Building in Ten Books.* London: MIT Press.
2. Alberti LB (2000) *Descriptio Urbis Romae (Vol. 56).* Switzerland: Librairie Droz.

3. Alexander C (2002) *The Nature of Order: The Process of Creating Life.* Berkeley: Taylor & Francis.

4. Anderson C (2008) The end of theory: the data deluge makes the scientific method obsolete. *Wired Magazine*, 23 June, 08.

5. Aureli PV (2011) *The Possibility of an Absolute Architecture.* London: MIT Press.

6. Banham R, Barker P, Hall P et al. (1969) Non-plan-pxperiment in freedom. *New Society* 13(338): 435–441.

7. Banham R (1980) *Theory and Design in the First Machine Age.* Cambridge, MA: MIT Press.

8. Barnes TJ and Wilson MW (2014) Big data, social physics, and spatial analysis: the early years. *Big Data & Society* 1(1): 1–14.

9. Batty M (2007) Model cities. *Town Planning Review* 78(2): 125–151.

10. Batty M (2009) *Urban Modeling. International Encyclopedia of Human Geography.* Oxford: Elsevier.

11. Batty M (2015) Models again: their role in planning and prediction. *Environment and Planning B: Planning and Design* 42(2): 191–194.

12. Bellman R (1961) *Adaptive Control Processes: A Guided Tour (Vol. 4).* Princeton: Princeton University Press.

13. Bettencourt LMA, Lobo J, Helbing D et al. (2007) Growth, innovation, scaling, and the pace of life in cities. *Proceedings of the National Academy of Sciences* 104(17): 7301–7306.

14. Branzi A (2006) *No-Stop City, Archizoom Associati*. Orléans, France: Hyx.

15. Brautigan R (1967) *All Watched Over by Machines of Loving Grace.* San Francisco, CA: Communication Company.

16. Brenner, N., & Schmid, C. (2015). *Towards a new epistemology of the urban?*.City, 19(2-3), 151-182.

17. Brin S and Page L (1998) The anatomy of a large-scale hypertextual web search engine. *Computer Networks and ISDN Systems* 30(1–7): 107–117.

18. Bühlmann V and Wiedmer M. (2008) *Pre-Specifics: Some Comparatistic Investigations on Research in Design and Art*. Zürich: JRP Ringier.

19. Bühlmann, Vera, Ludger Hovestadt, and Vahid Moosavi. "Coding as literacy." (2015).

20. Castells, M (2000) Grassrooting the space of flows, in cities in the telecommunications age. In: Wheeler JO, Aoyama Y and Warf B (eds) *The Fracturing of Geographies*. London: Routledge, pp.19.

21. Conway J. (1970) The game of life. *Scientific American* 223(4): 4.
22. Cranshaw J, Schwartz R, Hong JI et al. (2012) The livehoods project: utilizing social media to understand the dynamics of a city. In: *Proceedings of the Sixth International AAAI Conference on Weblogs and Social Media*, Pittsburgh, PA, pp. 58–65.
23. Curd P. (1996) *A Presocratics Reader: Selected Fragments and Testimonia.* Indianapolis: Hackett Publishing Company.
24. Curd P and McKirahan RD (2011) *A Presocratics Reader: Selected Fragments and Testimonia.* Indianapolis: Hackett Publishing Company.
25. Deleuze G (1993) *The Fold: Leibniz and the Baroque.* Minneapolis: University of Minnesota Press.
26. Eagle N and Pentland A (2006) Reality mining: sensing Complex social systems. *Personal and Ubiquitous Computing* 10(4): 255–268.
27. Eliade M (1954) *The Myth of the Eternal Return: Or, Cosmos and History.* Princeton: Princeton University Press.
28. Ellul J (1970) *The Meaning of the City.* Grand Rapids, MI: Eerdmans.
29. Evans D (2011) *Introduction to Computing: Explorations in Language, Logic, and Machines.* Charlottesville: University of Virginia.
30. Forrester JW (1969) *Urban Dynamics.* Cambridge, MA: MIT Press.
31. Foucault M (1970) *The Order of Things.* New York: Vintage.
32. Foucault M (1977a) A Preface to Transgression. *Language, Counter-Memory, Practice: Selected Essays and Interviews.* Blackwell, 29–52.
33. Foucault M (1977b) *Discipline and Punish: The Birth of the Prison.* New York: Vintage Books.
34. Foucault M (1980b) The confession of the flesh. In: Gordon C (ed) *Power/Knowledge Selected Interviews and Other Writings*, Brighton: Harvester Press, pp.194–228.
35. Foucault M (1979) *The History of Sexuality.* London: Allen Lane.
36. Foucault M (2002) *The Order of Things: An Archaeology of the Human Sciences.* London: Psychology Press.
37. Foucault M (2012) *The Archaeology of Knowledge.* New York: Vintage.
38. Friedman Y (2000) *Utopies Réalisables.* Paris: Éditions de l'Éclat.

39. Frigg R and Hartmann, S (2006) Models in science. In: Stanford Encyclopedia of Philosophy (Spring 2006 Edition).
40. Gleick J (2011) *The Information: A History, a Theory, a Flood.* New York: Pantheon Books.
41. Graham, S., & Marvin, S. (2001). *Splintering urbanism: networked infrastructures, technological mobilities and the urban condition.* Psychology Press.
42. Greenfield A (2013) *Against the Smart City. (The City is Here for You to Use)*. New York: Do Projects.
43. Hall PG (1988) *Cities of Tomorrow.* Oxford: Blackwell Publishers
44. Hansmann, U (Ed.) (2003). *Pervasive Computing*: *The Mobile World*. Heidelberg: Springer.
45. Haque U (2007) The architectural relevance of Gordon Pask. *Architectural Design* 77(4): 54–61.
46. Harris B (1964) *A Model of Locational Equilibrium for Retail Trade.* Philadelphia: Penn-Jersey Transportation Study.
47. Hey T, Tansley S and Tolle KM (2009) Jim grey on escience: a transformed scientific method. In: Hey T, S T and K T (eds) *The Fourth Paradigm: Data-Intensive Scientific Discovery.* Redmond: Microsoft Research, pp.xvii–xxxi.
48. Hillier B and Hanson J (1984) *The Social Logic of Space.* Cambridge: Cambridge University Press.
49. Hillier B (1996) *Space Is the Machine: A Configurational Theory of Architecture.* Cambridge: Cambridge University Press.
50. Hovestadt L and Buhlmann V (2014) *Eigen Architecture.* Ambra. Bilingual edition (January 2014)
51. Hovestadt L, Bühlmann V, Alvarez-Marin D et al. (2015) *A Quantum City: Mastering the Generic.* Vienna: Birkhäuser.
52. Jacobs J (1961) *The Death and Life of Great American Cities.* New York: Vintage Press.
53. Koolhaas R (1995) *Generic City.* Sassenheim: Sikkens Foundation.
54. Koolhaas R (2006) *Junkspace.* Macerata, Italy: Quodlibet
55. Kostof S (1991) *The City Shaped: Urban Patterns and Meanings through History.* London: Thames and Hudson.
56. Kuhn TS (1977) *The Essential Tension: Selected Studies in Scientific Tradition and Change.* Chicago: University of Chicago Press.

57. Lee DB (1973) Requiem for large-scale models. *Journal of the American Institute of Planners* 39(3): 163–178.

58. Lowry IS (1965) A short course in model design. *Journal of the American Institute of Planners* 31(2): 158–166.

59. Lynch K (1984) *A Theory of Good City Form.* Cambridge, MA: MIT Press.

60. Moosavi, V. (2015). *Pre-Specific Modeling* (Doctoral dissertation, Diss., Eidgenössische Technische Hochschule ETH Zürich, Nr. 22683).

61. Mumford L and Copeland G (1961) *The City in History: Its Origins, Its Transformations, and Its Prospects,* New York: Harcourt, Brace & World, pp.602615.

62. Nieuwenhuys C (1974) *New Babylon.* Constant: New Babylon, pp.154.

63. O'reilly T (2007) What is web 2.0: design patterns and business models for the next generation of software. *Communications & Strategies* 1: 17.

64. Ousterhout JK (1998) Scripting: higher level programming for the 21st century. *IEEE Computer Magazine* 31: 23–30.

65. Pasquinelli M (2009) *Google's pagerank algorithm: a diagram of cognitive capitalism and the rentier of the common intellect.* Deep Search 3: 152–162.

66. Paulos, E., Honicky, R., & Hooker, B. (2008). *Citizen science: Enabling participatory urbanism.* Urban Informatics: Community Integration and Implementation.

67. Pentland A (2014) *Social Physics: How Good Ideas Spread — the Lessons from a New Science.* New York: Penguin.

68. Pope A (1996) *Ladders.* Houston, TX: Rice School of Architecture.

69. Ratti, C., & Townsend, A. (2011). *The social nexus.* Scientific American, 305(3), 42-48.

70. Ricardo D and Li Q (1819) *The Principles of Political Economy and Taxation.* London: J. Murray.

71. Rosen G (2014) Abstract Objects. In: Zalta EN (ed) *The Stanford Encyclopedia of Philosophy.* Fall 2014 ed. URL: http://plato.stanford.edu/archives/fall2014/entries/abstract-objects/.

72. Rowe C and Koetter F (1983) *Collage City.* Cambridge, MA: MIT Press.

73. Rykwert J (1988) *The Idea of a Town: The Anthropology of Urban Form in Rome, Italy and the Ancient World.* Cambridge, MA: MIT Press.

74. Salingaros NA (2005) *Principles of Urban Structure (Vol. 4).* Delft: Techne Press.

75. Scott ML (2000) *Programming Language Pragmatics*. San Francisco: Morgan Kaufmann.

76. Shannon, C. E., & Weaver, W. (1949). *The mathematical theory of communication.* Urbana, IL.

77. Serres M (2013) Available at: www.academie-francaise.fr/actualites/communication-de-m-michel-serres.

78. Shane DG (2005) *Recombinant Urbanism: Conceptual Modeling in Architecture, Urban Design, and City Theory*. Chichester: Wiley.

79. Smithson AM (1957) Cluster city, a new shape for the community. *The Architectural Review* 122(730): 333–336.

80. Sorokin PA (1928) Contemporary Sociological Theories. New York: Harper.

81. Tobler W (1979) Cellular geography. In: Gale S and D GO (eds) *Philosophy in Geography.* Boston: Reidel, pp.379–386.

82. Townsend AM (2013) *Smart Cities: Big Data, Civic Hackers, and the Quest for a New Utopia.* New York: WW Norton & Company.

83. Turing AM (1936) On computable numbers, with an application to the Entscheidungsproblem. *Jounal of Mathematics* 58(345-363): 5.

84. United Nations, Department of Economic and Social Affairs, Population Division (2014). *World Urbanization Prospects: The 2014 Revision, Highlights* (ST/ESA/SER.A/352).

85. Venturi R (1977) *Complexity and Contradiction in Architecture (Vol. 1)*. New York: Museum of Modern Art.

86. Von Thünen JH and Schumacher-Zarchlin H (1875) Der Isolirte Staat. In: *Beziehung auf Landwirtschaft und Nationalökonomie (Vol. 1)*. Berlin: Wiegant Hempel & Parey.

*87.* Von Thünen JH (1826) Der Isolierte Staat.In *Beziehung auf Landwirtschaft und Nationalökonomie.*1 Aufl. 1826 Stuttgart.

88. Waddell P and Ulfarsson G (2004) Introduction to urban simulation: design and development of operational models. In: Hensher D, K Button, K Haynes and P Stopher (ed) *Handbook of Transport Geography and Spatial Systems, Volume 5 (Handbooks in Transport)*. Oxford, UK: Pergamon Press, pp.203–236.

89. Wassermann K (2011) 'Sema città. deriving elements for an applicable city theory', *In Proceedings of 29th eCAADe, Respecting Fragile Places.* Slovenia: University of Ljubljana, pp.134–142.

90. Weber A and Pick G (1909) *Über Den Standort Der Industrien.* Tübingen: JCB Mohr.

91. Weiser M (1991) The computer for the 21st century. *Scientific American* 265(3): 94–104.

92. Weiser M (1994) Ubiquitous computing. *ACM Conference on Computer Science.* 418.

93. Wiener N (1961) *Cybernetics or Control and Communication in the Animal and the Machine (Vol. 25).* Cambridge, MA: MIT Press.

94. Wilson A (2012) *Urban Modelling: Critical Concepts in Urban Studies (Volumes 1-5).* London: Routledge.

95. Wolman A (1965) The metabolism of cities. *Scientific American* 213(3): 179–190.

96. Yuan J, Zheng Y, Zhang C et al. (2010) T-drive: driving directions based on taxi trajectories. In: *Proceedings of the 18th SIGSPATIAL International Conference on Advances in Geographic Information Systems,* New York: ACM, pp.99–108.